\documentclass[preprint]{aastex} 
\usepackage{aastexug}  
\begin{document} 
\shorttitle{Timing observations of 17 pulsars}
\title{Arecibo timing observations of 17 pulsars along the Galactic plane}

\author{Duncan R. Lorimer}
\affil{Arecibo Observatory, HC3 Box 53995, Arecibo, PR~00612}
\affil{The University of Manchester, Jodrell Bank Observatory,
Macclesfield, Cheshire, SK11~9DL, UK}
\author{Fernando Camilo}
\affil{Columbia Astrophysics Laboratory, Columbia University, 550 West
120th Street, New York, NY~10027}
\and
\author{Kiriaki M. Xilouris}
\affil{Department of Astronomy, University of Virginia,
Charlottesville, VA~22903}

\begin{abstract} \noindent
We present phase-coherent timing solutions obtained for the first time for
17 pulsars discovered at Arecibo by Hulse \& Taylor (1975ab) in a 430-MHz
survey of the Galactic plane.  This survey remains the most sensitive of the
Galactic plane at 430\,MHz and has comparable equivalent sensitivity to the
1400-MHz Parkes multibeam survey.  Comparing both surveys we find that, as
expected, the one at 430\,MHz is limited in depth by interstellar dispersion
and scattering effects; and that the detection rate of pulsars with
high spin-down luminosity ($\dot E>10^{34}$\,erg\,s$^{-1}$) at the low
frequency is a factor of 5 smaller than at high frequency. We also present
scatter-broadening measurements for two pulsars and pulse nulling and
mode-changing properties for two others.
\end{abstract}


\section{Introduction} \label{sec:intro}

The first systematic search for pulsars using the 305-m Arecibo radio
telescope was carried out by Hulse \& Taylor (hereafter HT) over 25 years
ago \nocite{ht74,ht75a,ht75b} (HT 1974, 1975a, b).  The HT survey was
conducted at 430\,MHz and covered an area of about 140 square degrees in the
region defined by $42\arcdeg \la l \la 60\arcdeg$ and
$\mid\!\!b\!\!\mid\:\la 4\arcdeg$. An integration time of 136.5\,s resulted
in a limiting flux density to long-period pulsars of about 1\,mJy ---
roughly an order of magnitude more sensitive than any other pulsar survey at
that time, and still the most sensitive low-frequency survey along the Galactic
plane. A total of 40 pulsars were discovered, the most notable being the
double neutron star binary system B1913+16 \nocite{ht75a,tw82,tw89} (HT
1975a) which has since been used as a magnificent laboratory for
gravitational physics (Taylor \& Weisberg 1982, 1989).

Pulse period, position and dispersion measure (DM) determinations for the 39
long-period pulsars were published shortly after the completion of the
survey (HT 1975b). However, perhaps as a result of the interest in
observations of B1913+16, timing observations have been carried out for only
22 sources \nocite{gr78} (see e.g.~Gullahorn \& Rankin 1978).  In this paper
we report on new Arecibo timing observations of the remaining 17 pulsars
which have resulted in accurate parameters for all of them. In \S~2 we
describe the observations and procedures used to obtain the timing
solutions. The results are presented in \S~3. We discuss scattering and
single-pulse properties of the pulsars respectively in \S~4.1 and \S~4.2.
Finally, in \S~4.3, we briefly discuss the population properties of the
sample and compare them with pulsars discovered in the 1374-MHz Parkes
multibeam survey.

\section{Observations and Analysis} \label{sec:obs}

The observations were carried out using primarily the 430-MHz line-feed
receiver on the Arecibo telescope between 1999 May and 2000 November. During
each session we observed all 17 pulsars, as well as the strong pulsar
B1933+16 for control purposes.  The incoming 430-MHz signals from both
senses of circular polarization were amplified and down-converted to an
intermediate frequency of 30\,MHz.  The Penn State Pulsar Machine (PSPM), a
$2\times128\times60$-kHz filterbank spectrometer, was used to record the
data in one of two modes.  In ``search mode'', data are continuously sampled
every $80\,\mu$s with 4-bit precision for each of the 128 frequency channels. 
Off-line processing proceeds by first producing a dedispersed
time series in which successive channels are delayed in time by an amount
corresponding to the nominal DM of each pulsar. The time series can then
either be searched for periodic signals, or folded modulo a particular pulse
period to produce an integrated pulse profile.  In ``timing mode'', a
custom-built chip is used to fold the incoming data for each frequency
channel online. The resulting 128 profiles are then de-dispersed and summed
to produce a single profile. In both operating modes, the PSPM is
synchronized to start on a well-calibrated 10-s tick pulse so that all data
can be used for timing purposes.

Since all of the existing ephemerides for these pulsars dated back to the
original discovery papers published over 25 years ago, they could not be
used to predict the topocentric period with sufficient accuracy to fold the
data on-line. To obtain useful ephemerides, the first few sessions were used
to make observations of each pulsar in search mode.  From an analysis of
these data, we obtained an up-to-date measurement of the period of each
pulsar and verified that it was the fundamental period. Although good
positional determinations were available for several pulsars as a result of
previous observations \nocite{wbf+81,vms83} (Weisberg et al.~1981;
Vivekanand, Mohanty \& Salter 1983), an improvement was necessary for
several pulsars whose positions still had uncertainties on the order of the
half-power width of the telescope beam ($\pm 5$ arcmin for the HT
survey). Positions for these pulsars were improved by taking search-mode
data for a number of offset telescope pointings about the nominal position
and localizing the telescope pointing which gave the highest signal-to-noise
ratio.  The resulting period and position measurements were used to form
ephemerides which were adopted in all subsequent observing sessions which
used the PSPM directly in timing mode.  Integration times ranged between 90
and 180 s, with weaker pulsars demanding longer integrations to obtain
sufficient signal-to-noise ratio.

The analysis of the timing-mode profiles proceeded by forming a high
signal-to-noise template profile for each pulsar. This was done in the first
instance by shifting individual profiles so that the peak amplitude is at
phase 0.5 before adding a number of them (typically 10 profiles)
together. The time of arrival (TOA) of each profile was then calculated by
convolution with the template and adding the resulting time offset to the
time stamp recorded at the start of the observation. For further details of
this procedure, see Taylor (1992)\nocite{tay92}.  Based on this set of TOAs
and the initial ephemeris used for the timing observations, we used the
{\sc tempo}\footnote{\url{http://pulsar.princeton.edu/tempo}} 
software package to fit a simple spin-down model consisting of four free
parameters: pulse period, period derivative, right ascension and
declination.  The resulting ephemerides were then used to phase-align each
observed profile to the midpoint before adding to produce a better template
for each pulsar.  This new template was then used to generate new TOAs which
were in turn run through {\sc tempo} to obtain a better ephemeris. This
process significantly reduced the rms observed-minus-model TOA residuals,
increasing the precision of the resultant parameter estimates.

Previously published DM measurements of the 17 pulsars often have
uncertainties of the order $\pm 20$\,cm$^{-3}$\,pc.  In order to make more
precise DM measurements, in 2001 May we carried out multi-frequency timing
observations at 1175 and 1475\,MHz using the L-wide Gregorian receiver. Data
for these observations were collected with the Wide-band Arecibo Pulsar
Processor (WAPP) --- a fast-dump digital correlator which samples a 100-MHz
bandwidth \nocite{dsh00} (Dowd et al.~2000).  Off-line dedispersion and
folding of the WAPP data using the ephemerides derived above and standard
data processing tools (Lorimer \nocite{lor01b} 2001) produced time-tagged
integrated pulse profiles. Combining the resulting TOAs with the 430-MHz
data, a fit for DM using {\sc tempo} and holding all other parameters in the
ephemeris constant yielded DM uncertainties of the order of a few
cm$^{-3}$\,pc, or less. We verified the validity of our analysis by fitting
the 430-MHz TOAs separately with the 1175 and 1475-MHz data, which gave
consistent results. We note also that our DM determination for J2027+2146 is
in excellent agreement with the \nocite{han87} Hankins' (1987) measurement
($96.8\pm0.2$\,cm$^{-3}$\,pc). As a final iteration, the new DM values were
used to dedisperse the PSPM timing-mode data to produce definitive 430-MHz
template profiles for all the pulsars. For several pulsars the improved DM
resulted in a much sharper profile showing features that were previously
unresolved due to dedispersion at the less precise DM value.

Due to the 4-bit quantization scheme used in the PSPM, the averaged
pulse profiles obtained in timing mode range between 0 and 15 in each
bin.  The scaling factor to convert these profiles into flux density
units is $1000 \times T/(D \times G)$, where $T$ is the total system
noise temperate (K), $D$ is the DC offset of the profile (in PSPM
counts), and $G$ is the antenna gain (K\,Jy$^{-1}$). With this choice
of units, we obtain a profile in mJy. For Arecibo observations, both
$T$ and $G$ are strong functions of the zenith angle, $z$, as both the
illumination pattern and system noise temperature change from their
optimum values at the zenith. To include these effects, we used the
analytical expressions for $T(z)$ and $G(z)$ determined by Perillat
(1999) \nocite{per99b} from continuum source calibration
observations. The adopted system temperature also includes the
contribution from the sky background radiation obtained from the
all-sky survey of \nocite{hks+82} Haslam et al.~(1982).

\begin{deluxetable}{lllllrcccc}
\tabletypesize{\scriptsize}
\tablewidth{0pt}
\tablecaption{\label{tab:obsparms}Observed parameters of 17 pulsars}
\renewcommand{\tabcolsep}{1.5mm}
\tablecolumns{10}
\tablehead{
\colhead{PSR}                  &
\colhead{R. A.}                &
\colhead{Dec.}                 &
\colhead{$P$\tablenotemark} &
\colhead{$\dot P$}             &
\colhead{DM}                   &
\colhead{$w_e$}                &
\colhead{$w_{50}$}             &
\colhead{$w_{10}$}             &
\colhead{$S_{430}$}           \\
\colhead{}                     &
\colhead{(J2000)}              &
\colhead{(J2000)}              &
\colhead{(s)}                  &
\colhead{$(10^{-15})$}         &
\colhead{($\rm{cm^{-3}\,pc}$)} &
\colhead{(ms)}                 &
\colhead{(ms)}                 &
\colhead{(ms)}                 &
\colhead{(mJy)}                }
\startdata
J1904+1011 & 19 04 02.49(2)  & +10 11 34.6(5)  & 1.85656971650(7) 
 &  0.276(8)  & 135(2) & 51 & 28;37 & 84;93 & 4.4(3) \\
J1907+1247 & 19 07 10.70(2)  & +12 47 35.9(5)  & 0.82709737059(3)
 &  1.948(4)  & 257(1) & 19 & 20 & 33 & 0.8(1) \\
J1912+1036 & 19 12 46.33(1)  & +10 36 41.6(3)  & 0.409349485862(6) 
 & 15.7620(8) & 147.0(5) & 16 & 15 & 31 & 1.6(1) \\
J1913+0936 & 19 13 52.70(2)  & +09 36 41.8(3)  & 1.24196459936(3) 
 &  0.432(4)  & 157(2) & 29 & 18 & 51 & 0.8(1) \\
J1921+2003 & 19 21 51.597(8) & +20 03 20.8(8)  & 0.76068138902(1) 
 &  0.050(1)  &  101(1) & 13 & 8;11  & 15;27 & 2.3(1) \\
J1923+1706 & 19 23 07.85(1)  & +17 06 09.4(2)  & 0.547209201831(8) 
 &  0.043(1)  & 142.5(6) & 16 & 19 & 28 & 1.5(1) \\
J1926+1928 & 19 26 22.82(4)  & +19 28 11.7(8)  & 1.34601218493(9)
 &  1.43(1)   & 445(2) & 47 & 54 & 88 & 0.8(1) \\
J1927+1852 & 19 27 10.422(8)  & +18 52 08.5(2)  & 0.482766273821(7)
 &  0.116(1)  & 254(1) & 29 & 32 & 52 & 3.4(1) \\
J1927+1856 & 19 27 24.97(1)  & +18 56 36.8(2)  & 0.298313497249(4) 
 &  2.2430(8)   & 99(1) & 18 & 9 & 43 & 2.2(1) \\
J1929+1844 & 19 29 16.78(4)  & +18 44 59.5(7)  & 1.22047000453(8) 
 &  2.36(1)   & 112(2) & 25 & 25 & 49 & 1.7(2) \\
J1931+1536 & 19 31 55.71(1)  & +15 36 57.5(3)  & 0.314355396469(6) 
 &  5.0148(9) & 140(1) & 12 & 11 & 27 & 1.2(1) \\
J1933+1304 & 19 33 22.529(4) & +13 04 49.97(8) & 0.928323751417(5)
 &  0.3182(7) & 177.9(2) & 16 & 7  & 37 & 2.0(1) \\
J1935+1745 & 19 35 29.97(1)  & +17 45 12.2(2)  & 0.654408146583(1) 
 &  0.378(2)  & 214.6(4) & 27 & 24 & 60 & 1.3(1) \\
J1942+1743 & 19 42 01.03(2)  & +17 43 28.3(4)  & 0.69626221476(2) 
 &  0.101(3)  & 190(6) & 42 & 46 & 92 & 2.8(1) \\
J1944+1755 & 19 44 31.80(4)  & +17 55 42.4(7)  & 1.9968990135(1) 
 &  0.73(2)   & 175(2) & 50 & 40;20 & 85;70 & 1.9(1) \\
J1945+1834 & 19 45 36.10(2)  & +18 34 20.1(4)  & 1.06870769134(4)
 &  0.242(5)  & 217.7(1) & 28 & 28 & 49 & 1.2(1) \\
J2027+2146 & 20 27 16.690(3) & +21 46 04.43(5) & 0.398173021652(2)
 &  0.2028(2) &  96.8(4) & 9 & 6 & 21 & 0.7(1) \\
\enddata
\tablecomments{Epoch of period is 51600.0 (MJD) for all sources.  Units
of right ascension are hours, minutes, and seconds. Units of
declination are degrees, arcminutes, and arcseconds.  Figures in
parentheses are 1\,$\sigma$ uncertainties in the least-significant
digits.}
\end{deluxetable}

Using this calibration technique, phase-averaged flux densities were
obtained for each profile by simply calculating the area under each pulse
and dividing this by the total number of bins in the profile.  There was
often a significant (few arcmin) offset between the positions used to point
the telescope for the timing observations and the actual pulsar positions
obtained following the timing analysis.  As a result, our measured flux
densities underestimate the true values slightly. To account for this, each
measured flux density was multiplied by the correction factor
$\exp(4\ln2\times\Delta^2/W^2)$ where $\Delta$ is the offset from the center of
the assumed Gaussian telescope beam and $W$ is the half-power beamwidth
(10\,arcmin for the 430-MHz system).

\section{Results} \label{sec:res}

The timing parameters, pulse widths and flux densities are presented in
Table~\ref{tab:obsparms}. Listed are pulsar name, right ascension and
declination, pulse period $P$ and its derivative $\dot{P}$, DM, equivalent
pulse width $w_e$, pulse widths $w_{50}$ and $w_{10}$ measured respectively
at 50 and 10\% of the peak amplitude, and the mean 430-MHz phase-averaged
flux density $S_{430}$. Separate values of $w_{50}$ and $w_{10}$ are
tabulated for the three pulsars with double-pulse profiles.  The equivalent
width $w_e$ is the width of a top-hat pulse having the same area and peak
height as the profile.  $S_{430}$ is the mean of all the individual 
flux-density measurements. Using our data for PSR~B1933+16, we find
$S_{430}=233\pm6$\,mJy, in excellent agreement with the value of
$242\pm22$\,mJy based on 408-MHz Jodrell Bank \nocite{lylg95} observations
(Lorimer et al.~1995).  Post-fit timing residuals and 430-MHz
profiles\footnote{available online at the European
Pulsar Network profile database:
\url{http://www.mpifr-bonn.mpg.de/div/pulsar/epn}}  are shown in Figure~1. 

Derived parameters listed in Table~\ref{tab:derparms} are: distance $d$,
calculated using the Taylor \& Cordes (1993) \nocite{tc93} Galactic electron
density model; Galactic height $z = d \sin b$; 430-MHz luminosity $L_{430}
\equiv S_{430} d^2$; spin-down age $\tau = P/(2\dot{P})$; spin-down energy
loss rate $\dot{E} = 4\pi^2 I \dot P/P^3$ (where the moment of inertia $I$
is taken to be $10^{45}$\,g\,cm$^{2}$; Manchester \& Taylor
1977\nocite{mt77}); and dipole magnetic field strength $B=3.2\times10^{19}
(P\dot{P})^{1/2}$\,Gauss.

\begin{figure}[hbt]
\plotone{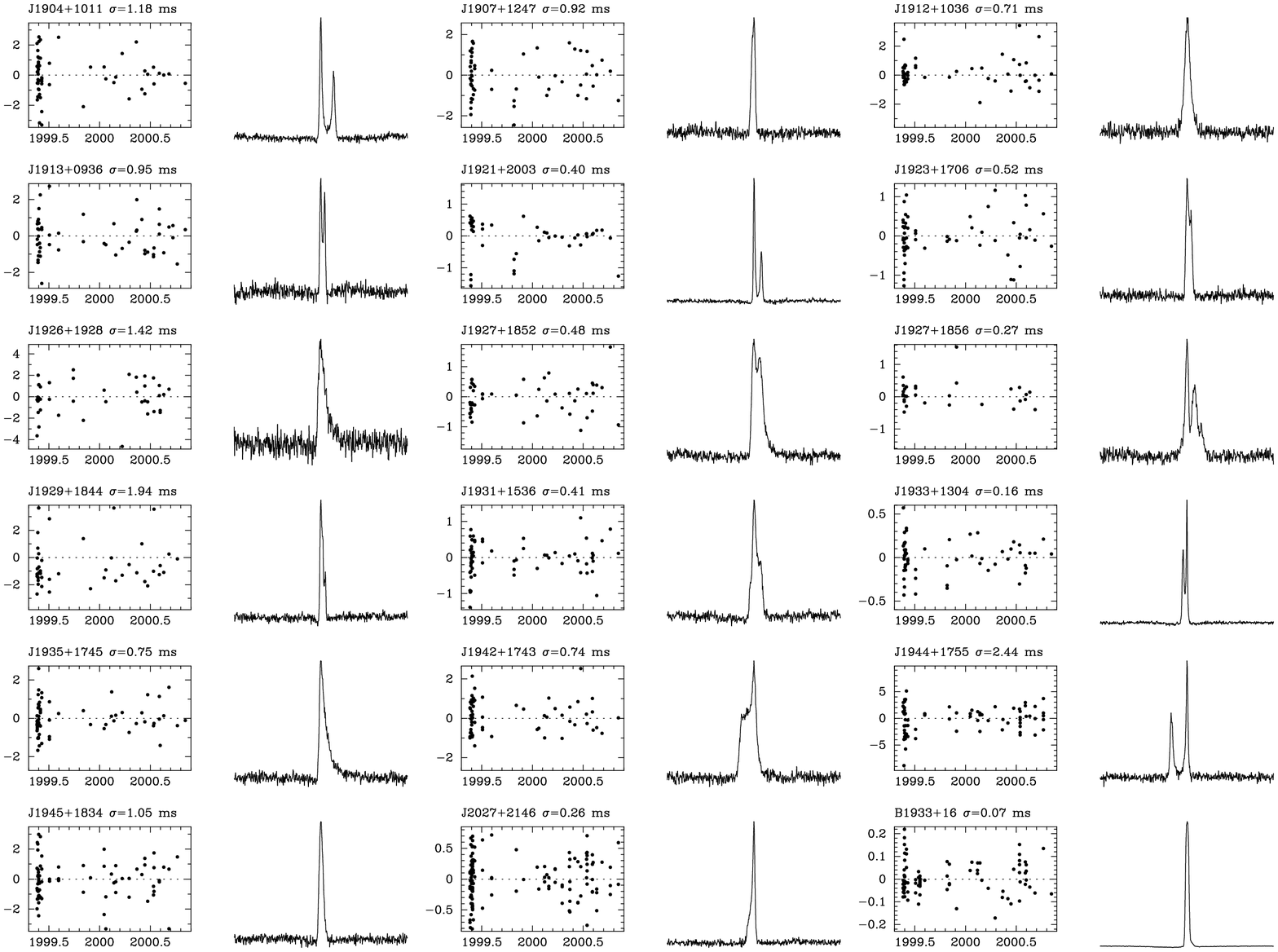}
\caption{
Post-fit timing residuals (in ms, with rms value given by $\sigma$) vs.
date, and integrated pulse profiles from 430-MHz PSPM observations of
18 pulsars (including the calibrator PSR~B1933+16).  Profile time
resolution is 2 milliperiods in all cases. }
\end{figure}

\begin{deluxetable}{lrrcccc}
\tabletypesize{\scriptsize}
\tablewidth{0pt}
\tablecaption{\label{tab:derparms} Derived parameters of 17 pulsars}
\tablecolumns{7}
\tablehead{
\colhead{PSR}                              &
\colhead{$d$} 			           &
\colhead{$z$}                              &
\colhead{$\log L_{430}$}                   &
\colhead{$\log \tau$}		  	   &
\colhead{$\log \dot E$}			   &
\colhead{$\log B$}		 	 \\
\colhead{}                &
\colhead{(kpc)}           &
\colhead{(kpc)}           &
\colhead{(mJy\,kpc$^2$)}  &
\colhead{(yr)}            &
\colhead{(erg\,s$^{-1}$)} &
\colhead{(G)}          }
\startdata
J1904+1011 &  4.0 &   0.13  & 1.9 & 8.03 & 30.23 & 11.86 \\
J1907+1247 &  7.2 &   0.30  & 1.6 & 6.83 & 32.13 & 12.11 \\
J1912+1036 &  4.2 &   0.01  & 1.5 & 5.61 & 33.96 & 12.41 \\
J1913+0936 &  4.2 & $-0.04$ & 1.1 & 7.66 & 30.95 & 11.87 \\
J1921+2003 &  5.2 &   0.24  & 1.8 & 8.38 & 30.66 & 11.30 \\
J1923+1706 &  6.4 &   0.11  & 1.8 & 8.30 & 31.02 & 11.19 \\
J1926+1928 & 30   &   0.74  & 2.9 & 7.17 & 31.36 & 12.15 \\
J1927+1852 & 10   &   0.17  & 2.6 & 7.82 & 31.61 & 11.38 \\
J1927+1856 &  4.8 &   0.08  & 1.7 & 6.32 & 33.52 & 11.92 \\
J1929+1844 &  5.3 &   0.04  & 1.7 & 6.91 & 31.71 & 12.23 \\
J1931+1536 &  6.4 & $-0.18$ & 1.7 & 6.00 & 33.80 & 12.10 \\
J1933+1304 &  8.1 & $-0.44$ & 2.1 & 7.66 & 31.20 & 11.74 \\
J1935+1745 &  9.6 & $-0.22$ & 2.1 & 7.44 & 31.73 & 11.70 \\
J1942+1743 &  9.8 & $-0.46$ & 2.4 & 8.04 & 31.07 & 11.43 \\
J1944+1755 &  9.5 & $-0.51$ & 2.2 & 7.64 & 30.56 & 12.09 \\
J1945+1834 & 13   & $-0.67$ & 2.3 & 7.84 & 30.89 & 11.71 \\
J2027+2146 & 10   & $-1.68$ & 1.8 & 7.49 & 32.10 & 11.46 \\
\enddata
\end{deluxetable}

\section{Discussion} \label{sec:disc}

\subsection{Scatter Broadening of Pulse Profiles} \label{sec:scat}

The profiles for PSRs~J1926+1928 and J1935+1745 (Figure~1) show clear
examples of ``scattering tails'', the excess delay caused by multi-path
scattering from irregularities in the electron density of the interstellar
medium \nocite{sch68} (Scheuer 1968).  Different path lengths between
scattered and unscattered rays result in a range of arrival times which
broadens intrinsically sharp pulses in an exponential fashion.  Fitting the
observed profiles to a one-sided exponential function with a time constant
$\tau_s$ using the procedure described by L\"ohmer et al.~(2001),
\nocite{lkm+01} we find $\tau_s$ values of $47\pm6$\,ms and $24\pm1$\,ms for
PSRs~J1926+1928 and J1935+1745, respectively, at 430\,MHz.

In general, the amount of scattering increases as the number of scattering
electrons along the line of sight increases.  From published measurements,
Cordes (2001) \nocite{cor01} finds the following empirical relationship:
\begin{displaymath}
\log \tau_s = -3.59+0.129\log\rm{DM}+1.02(\log\rm{DM})^2-4.4\log\nu,
\end{displaymath}
where the units are $\mu$s, cm$^{-3}$\,pc and GHz for $\tau_s$, DM and
observing frequency $\nu$, respectively. Note the use of the Kolmogorov
power-law dependence $\tau_s \propto \nu^{-4.4}$ (see \nocite{ric77}
e.g.~Rickett 1977) for frequency scaling in this expression. As noted by
Cordes (2001) there is considerable variance about this best-fit expression
for individual pulsars. This can be seen for PSRs~J1926+1928 and J1935+1745
where the predicted values (329 and 7\,ms respectively) differ significantly
from our measurements.  These differences reflect the variation in
the clumpiness of the interstellar medium along different lines of sight, as
well as the deviation from the theoretical power-law dependence on observing
frequency (L\"ohmer et al.~2001).

Two other pulsars in Figure~1, J1907+1247 and J1927+1852, have predicted
430-MHz $\tau_s$ values of order 20\,ms.  Clearly no significant scattering
tail is evident in the profile for J1907+1247.  The profile for J1927+1852
appears to be possibly affected by scattering. However, the truncated
exponential fitting procedure used above does not result in a good fit to
this profile. A profile component analysis, using the procedure described by
\nocite{kra94} Kramer (1994), shows that the profile is best described by a
simple three-component Gaussian model rather than any exponentially decaying
components.

\begin{figure}[hbt]
\plotone{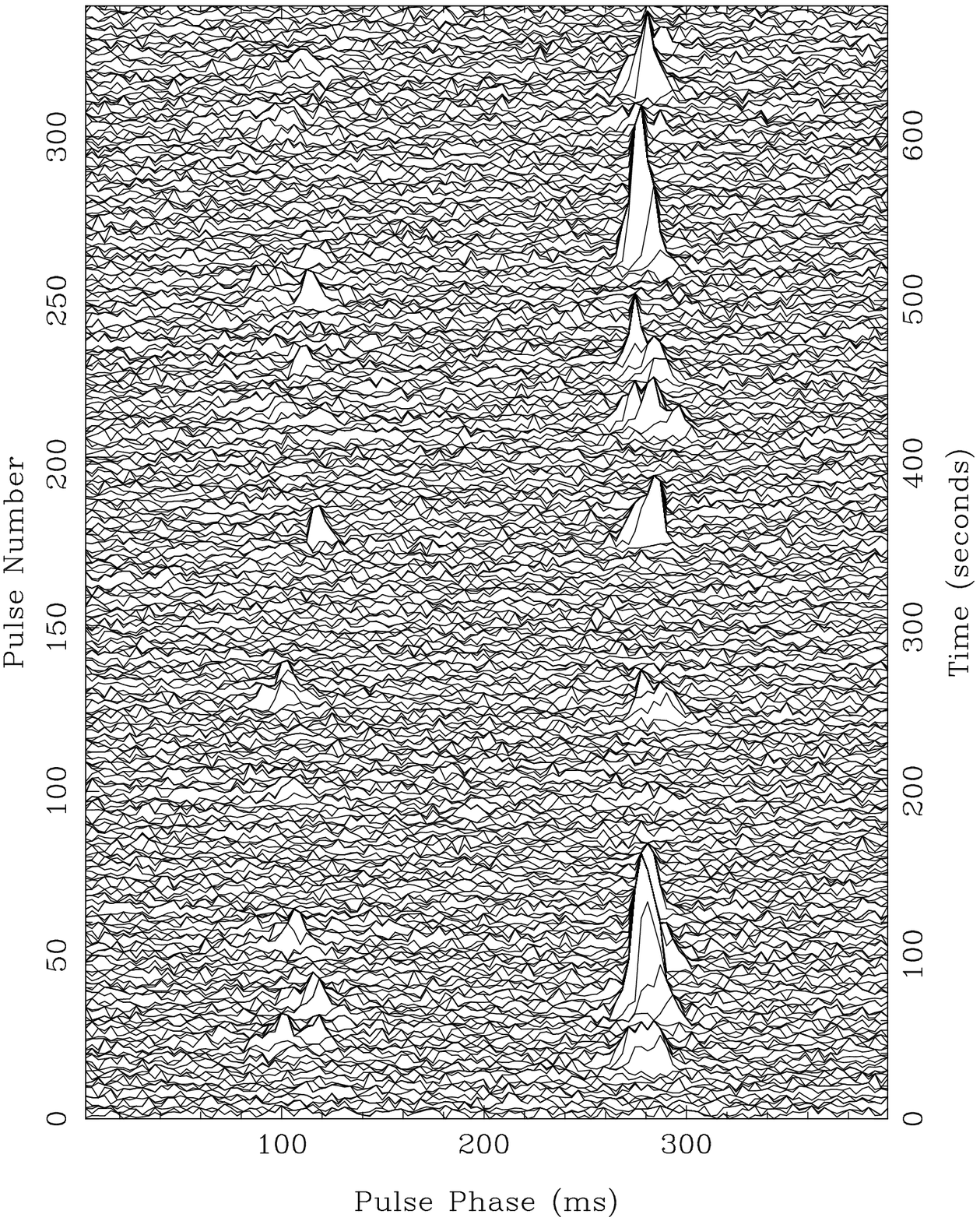}
\caption{A sequence of 340 consecutive single pulses
from the 2-s pulsar J1944+1745. For clarity, only the central 400\,ms of pulse
phase of each pulse is shown.}
\end{figure}

\subsection{Single-Pulse Properties} \label{sec:sp}

After obtaining the timing solutions we dedispersed and folded the original
search-mode PSPM data to examine the individual pulse properties.  It was
notable that the two longest period pulsars in the sample, PSRs~J1944+1755
and J1904+1011 $(P \sim 2$\,s), showed clear evidence for pulse-nulling: the
abrupt switch-off of emission for many pulse periods (Backer \nocite{bac70a}
1970a). A sequence of 340 consecutive single pulses from J1944+1755 is shown
in Figure~2. From a visual inspection of these data, we estimate that the
nulling fraction is at least 60\%. This is comparable with nulling observed
in other long-period pulsars \nocite{rit76} (Ritchings 1976).  As can be
seen in Figure~2, the pulsar often spends 1--2\,min in a null.  As a result,
this pulsar would not have been discovered had it been in this state during
the original observation of this position in the HT survey.  The number of
pulsars that have been missed by surveys with short integration times may be
large. A detailed analysis of the nulling characteristics of pulsars found
in the Parkes multibeam survey (Manchester et al.~2001), where much longer
35-min integration times are used, will be useful to quantify the population
of nulling pulsars.

The nulling fraction for PSR~J1904+1011 is harder to estimate since the
signal-to-noise ratios of the individual pulses were lower.  From the
timing-mode profiles it is clear that there is more than one stable state of
this profile.  Approximately 90\% of all timing-mode profiles collected were
similar to the one shown in Figure~1, where the leading component is about
twice the height of the second one. In the remaining 10\% of the profiles,
the second component was either of equal magnitude to, or stronger than, the
first one. It seems likely that this pulsar exhibits mode changing
\nocite{lyn71a,bac70b} (Backer 1970b; Lyne 1971).

The search-mode data we have collected are not ideal for detailed study of
the single-pulse behavior. Integration times were sufficient for collection
of only a few hundred pulses and the original telescope pointings were often
offset significantly from the true positions obtained from the timing
analysis. However, they do suggest that a dedicated program of single-pulse
observations with the new ephemerides presented here would be a worthwhile
addition to our existing knowledge of the nulling and mode-changing
phenomena in general \nocite{ran86} (Rankin 1986).

\subsection{Population Properties} \label{sec:pop}

For many years, the HT survey, with a limiting 430-MHz flux density of $\sim
1$\,mJy, provided the deepest sample of pulsars along the northern Galactic
plane. The on-going Parkes multibeam (hereafter PM) survey, with a limiting
flux density of 0.15\,mJy at a center frequency of 1374\,MHz, is now
providing a much larger sample of pulsars along much of the Galactic plane
\nocite{mlc+01} ($260\arcdeg < l < 50\arcdeg$ and $\mid\!\!b\!\!\mid\ <
5\arcdeg$; Manchester et al.~2001).  For a given pulsar spectral index
$\alpha$, the effective 430-MHz sensitivity of the PM survey is $0.15 \times
(430/1374)^{\alpha}$\,mJy.  For the mean pulsar spectral index value of
$-1.6$ (Lorimer et al.~1995)\nocite{lylg95} the nominal sensitivity limits
of the HT and PM surveys are similar.  Given this close match in
sensitivities, it is interesting to compare the properties of the pulsars
found in the two surveys.

HT (1975a, b) suggested that their search appeared to have detected a limit
to the spatial extent of pulsars in the search region ($42\arcdeg \la l \la
60\arcdeg$ and $\mid\!\!b\!\!\mid \, \la 4\arcdeg$). This idea was based on
the fact that only 2 out of the 49 pulsars detected in the HT survey had
$\mbox{DM} > 260$\,cm$^{-3}$\,pc. The median DM value in the sample is only
165\,cm$^{-3}$\,pc.  For the 24 pulsars published so far from the PM
survey\footnote{\url{http://www.atnf.csiro.au/research/pulsar/pmsurv/pmpsrs.db}}
which are visible from Arecibo we see a much broader distribution in DM with
a median of 347\,cm$^{-3}$\,pc and a maximum value of almost
800\,cm$^{-3}$\,pc. Rather than being a real effect, the DM cutoff seen in
the HT sample reflects the selection effect against detecting highly
scattered pulsars with large DMs in low-frequency surveys (see
\nocite{joh94} e.g.~Johnston 1994).

In the period--period derivative diagram presented in Figure~3 we compare
the spin properties of the HT pulsars with the currently available sample of
409 pulsars from the PM survey.  Also shown on this diagram is the locus of
points for which $\dot{E}=10^{34}$\,erg\,s$^{-1}$. The most striking
difference between the two samples is the lack of high $\dot{E}$ objects
found in the HT survey. Only 4\% of the HT pulsars have
$\dot{E}>10^{34}$\,erg\,s$^{-1}$ compared to 25\% of PM pulsars found in the
Arecibo declination range and 18\% for the PM survey as a whole. Since high
$\dot{E}$ pulsars are predominantly young, we can reasonably expect them to
be found close to their birth sites on the Galactic plane. This effect is
clearly seen in the sample of pulsars from the PM survey where almost 60\%
of the sample of objects with $\dot{E}>10^{34}$\,erg\,s$^{-1}$ are found at
latitudes in the range $\mid\!\!b\!\!\mid\:<0.5\arcdeg$.  Only 30\% of the
PM pulsars with $\dot{E}<10^{34}$\,erg\,s$^{-1}$ lie in the latitude range
$\mid\!\!b\!\!\mid\:<0.5\arcdeg$. At these low Galactic latitudes the high
sky background temperature on the plane, as well as pulse dispersion and
scattering, significantly reduce the sensitivity of low-frequency pulsar
surveys. These selection effects seem to be the most likely reason for the
dearth of high-$\dot{E}$ pulsars in the HT sample.

\begin{figure}[hbt]
\plotone{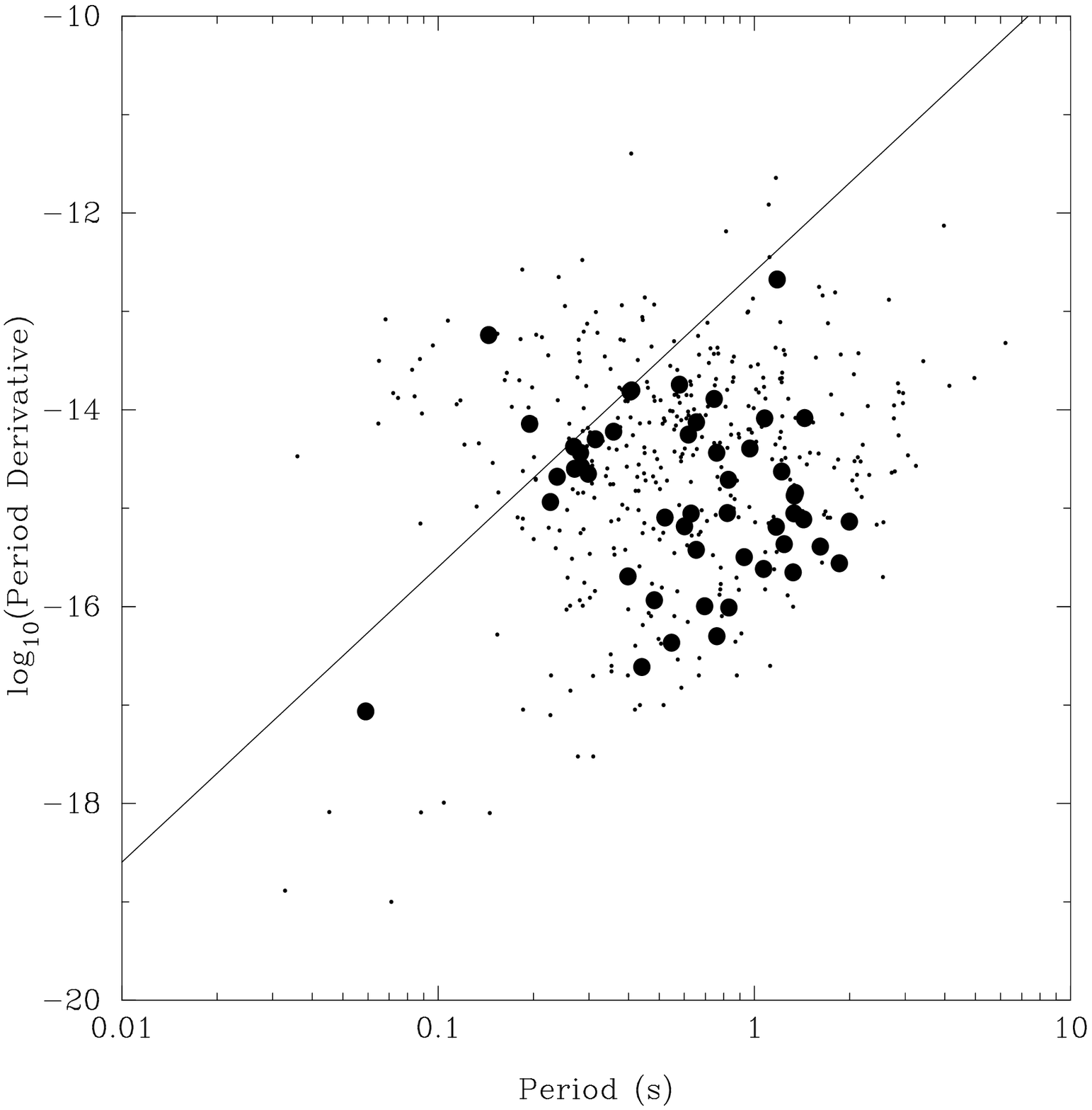}
\caption{
The $P$--$\dot P$ diagram.  Large points represent all pulsars detected
by the HT survey. Small points show the current sample of
409 pulsars from the PM survey. Pulsars to the left of the sloping line
have $\dot{E}>10^{34}$\,erg\,s$^{-1}$. The fraction of such objects
found in the PM survey is five times higher than for the HT survey. }
\end{figure}

The success of the PM survey in finding new pulsars in the Arecibo
declination range implies that many further discoveries could be made by a
large-scale Arecibo search using a 1400-MHz multibeam system.  Development
of such a system is currently underway and it is expected that it will be
available sometime in 2003 (P.~Goldsmith \& J.~Cordes, private
communication).  For integration times of about 300\,s per pointing and
bandwidths of about 200\,MHz, a survey with the Arecibo multibeam system
should be able to improve upon the sensitivity achieved by the PM survey by
a factor of about 5.  Although the pulsars observed in this study were the
weakest subset of the HT sample (with 430-MHz luminosities ranging between
10 and 1000\,mJy\,kpc$^2$; see Table~\ref{tab:derparms}), it is well known
that the pulsar luminosity function extends down to 1\,mJy\,kpc$^2$ at
430\,MHz and probably much fainter \nocite{lml+98} (Lyne et al.~1998). Thus,
in addition to finding many pulsars in general, a future Arecibo multibeam
survey would be an excellent probe of the low end of the luminosity function
for young pulsars.

\acknowledgments

\noindent
The Arecibo Observatory, a facility of the National Astronomy and
Ionosphere Center, is operated by Cornell University under a
cooperative agreement with the National Science Foundation.  We wish
to thank Alex Wolszczan for making the PSPM freely available for use
at Arecibo. Without this instrument, the observations presented here
would not have been possible. Thanks are also due to Ingrid Stairs for
providing software to calculate TOAs, Oliver L\"ohmer and Michael
Kramer for discussions and software to calculate scattering timescales
and Chris Salter, Dick Manchester and Maura McLaughlin for comments on
an earlier version of the manuscript. DRL is a University Research
Fellow funded by the Royal Society. FC is supported by NASA grant
NAG5-9095.

\end{document}